\documentclass[11pt]{article}

\usepackage{amsmath, latexsym, amssymb, amscd, amsfonts, amsthm, cite, epsfig, hyperref, enumerate, graphics, subfig}
\usepackage[numbers]{natbib}

\newcommand{\Tr}{\textrm{Tr}}							%Trace%
\newcommand{\be}{\begin{equation}}
\newcommand{\ee}{\end{equation}}

\begin{document}

\title{Asymptotic inference in system identification\\ for the atom maser}
\author{C\u{a}t\u{a}lin C\u{a}tan\u{a}, Merlijn van Horssen and M\u{a}d\u{a}lin Gu\c{t}\u{a}\\[4mm]
University of Nottingham, School of Mathematical Sciences\\ University Park, NG7 2RD Nottingham, UK}
\date{}

\maketitle
\begin{abstract}System identification is an integrant part of control theory and plays an increasing role in quantum engineering. In the quantum set-up, system identification is usually equated to process tomography, i.e. estimating a channel by probing it repeatedly with different input states. However for quantum dynamical systems like quantum Markov processes, it is more natural to consider the estimation based on continuous measurements of the output, with a given input which may 
be stationary. We address this problem using asymptotic statistics tools, for the specific example of estimating the Rabi frequency of an atom maser. We compute the Fisher information of different measurement processes as well as the quantum Fisher information of the atom maser, and establish the local asymptotic normality of these statistical models. 
The statistical notions can be expressed in terms of spectral properties of certain deformed Markov generators and the connection to large deviations is briefly discussed.

\end{abstract}
 
%\keywords{atom maser, quantum Markov processes, system identification, Fisher information, asymptotic normality} 
%\classification{...} 

%\begin{document}

%=========================================================%
\section{Introduction}
%=========================================================%

%We study the statistical properties of a physical system with dissipative dynamics known as the atom maser. We will introduce the mathematical description of this model and consider some of its features before moving on to our results.

%\subsection{Atom maser}

We are currently entering a new technological era \cite{Dowling&Milburn} where quantum control is a becoming a key component of quantum engineering  \cite{Mabuchi&Khaneja}. In the standard set-up of quantum filtering and control theory \cite{Belavkin1999,Wiseman&Milburn}  the dynamics of the system and its environment, as well as the initial state of the system, are usually assumed to be known. In practice however,  these objects may depend on unknown parameters and inaccurate models 
may compromise the control objective.  Therefore, system identification \cite{Ljung} which lies at the intersection of control theory and statistics, is becoming an increasingly relevant topic for quantum engineering \cite{Haffner}.

In this paper we introduce probabilistic and statistical tools aimed at a better understanding of the measurement process, and at solving system identification problems  in the set-up of quantum Markov processes. Although the mathematical techniques have a broader scope, we focus on the physically relevant model of the atom maser which has been extensively investigated both theoretically \cite{Briegel1994,Englert2002a} and experimentally \cite{Rempe,Walther} and found to exhibit a number of interesting dynamical phenomena. In the standard set-up of the atom maser, a beam of two-level atoms prepared in the excited state passes through a cavity with which they are in resonance, interact with the cavity field, and are detected after exiting the cavity. We consider the case where the atoms are measured in the standard basis, but more general measurements can be analysed in the same framework.

The specific questions we want to address are how to estimate the strength of the interaction between cavity and atoms, what is the accuracy of the estimator, and how it relates to the spectral properties of the Markov evolution. More generally, we aim at developing techniques for treating identifiability for multiple dynamical parameters, finding the associated Fisher information matrix, establishing asymptotic normality of the estimators. These topics are well understood in the classical context and our aim is to adapt and extend such techniques to the quantum set-up. In classical statistics it is known that if we observe the first $n$ steps of a Markov  chain whose transition matrix depends on some unknown parameter $\theta$, then $\theta$ can be estimated with an optimal asymptotic mean square error of $(n I (\theta))^{-1}$ where $I(\theta)$ is the Fisher information (per sample) of the Markov chain. Moreover the error is asymptotically normal (Gaussian)
\begin{equation}\label{eq.1}
\sqrt{n}(\hat{\theta}_{n} -\theta) \overset{\mathcal{L}}{\longrightarrow} N(0, I(\theta)^{-1}).
\end{equation}

The key feature of our estimation problem is that the atom maser's output consists of atoms which are correlated with each other and with the cavity. Therefore,  state and process tomography methods do not apply directly. In particular it is not clear what is the optimal measurement, what is the quantum Fisher information of the output, and how it compares with 
the Fisher information of simple (counting) measurements. These questions were partly answered in \cite{Guta} in the context of a discrete time quantum Markov chain, and here we extend the results to a continuous time set-up with an infinite dimensional system. For a better grasp of the statistical model we consider several thought and real experiments, and compute the Fisher informations of the data collected in these experiments. For example we analyse the set-up where the cavity is observed as it jumps between different Fock states when counting measurements are performed on the output as well as in the temperature bath. We also consider estimators which are based solely on the statistic given by the total number of ground or excited state atoms detected in a time period. Our findings are illustrated in Figures \ref{fig.Fisher} and \ref{fig.Fisher2} which shows the dependence of different (asymptotic)  Fisher informations on the interaction parameter. In particular the quantum Fisher information of the closed system made up of atoms, cavity and bath, depends strongly on the value of the interaction strength and is proportional to the cavity mean photon number in the stationary state. Furthermore, we find that monitoring all channels achieves the quantum Fisher information, while excluding the detection of excited atoms leads to a drastic decrease of the Fisher information in the neighbourhood of the first `transition point'. The asymptotic regime relevant for statistics is that of central limit, i.e. moderate deviations around the mean of order $n^{-1/2}$ as in \eqref{eq.1}. One of our  main results is to establish local asymptotic normality for the atom counting process, which implies the Central Limit and provides the formula of the Fisher information. We also prove a quantum version of this result showing that the quantum statistical model of the atom maser can be approximated in a statistical sense by a displaced coherent state model. The moderate deviations regime analysed as well as the related  regime of large deviations are closely connected to the spectral properties of certain deformations of the Lindblad operator. Some of these connections are pointed out in this paper 
but other questions such as the existence of dynamical phase transitions  \cite{Garrahan2009,Garrahan2011} and the quantum Perron-Frobenius Theorem will be addressed elsewhere \cite{Guta&vanHorssen}.

In sections \ref{sec.maser} and  \ref{sec.overview.statistics}  we give brief overviews of the atom maser's dynamics, 
and respectively of classical and quantum statistical concepts used in the paper. Section \ref{sec.Fisher} contains the main results about Fisher information and asymptotic normality in different set-ups. We conclude with comments on future work. 

%Another important feature is that the Fisher information contained in the total number of ground state atoms 

\section{The atom maser}\label{sec.maser}

The  atom maser's dynamics is based on the Jaynes-Cummings model of the atom-cavity Hamiltonian
\begin{equation}\label{eq.jaynes}
		H = H_{free} + H_{int}= \hbar\Omega a^{*} a + \hbar \omega \sigma^{*} \sigma
		- \hbar g(t) ( \sigma  a^{*} + \sigma^{*} a)
\end{equation}
where  $a$ is the annihilation operators of the cavity mode, $\sigma$ is the lowering operator of the two-level atom, 
$\Omega$ and $\omega$ are the cavity frequency and the atom transition frequency which are assumed to be equal, 
and $g$ is the coupling strength or Rabi frequency. In the standard experimental set-up the atoms prepared in the excited state arrive as a Poisson process of a given intensity, and interact with the cavity for a fixed time. Additionally, the cavity is in contact with a thermal bath with mean photon number $\nu$. 
By coarse graining the time evolution to ignore the very short time scale changes in the cavity field, one arrives at the following master equation for the cavity state $\rho$
 \begin{equation}
\frac{d\rho}{d t} = \mathcal{L} (\rho),
\end{equation} 
where $\mathcal{L}$ is the Lindblad generator
\begin{equation}\label{eq.lindblad}
		\mathcal{L}(\rho) = \sum_{i=1}^{4}\left( L_{i} \rho L_{i}^{*} - \frac{1}{2}\lbrace L_{i}^{*} L_{i}, \rho \rbrace \right);
\end{equation}
with operators
\begin{align} L_{1} &= \sqrt{N_{ex}} a^{*} \frac{\sin(\phi \sqrt{a a^{*}})}{\sqrt{a a^{*}}}, \quad L_{2} = \sqrt{N_{ex}} \cos(\phi \sqrt{a a^{*}}),\\
		L_{3} &= \sqrt{\nu+1}a, \quad L_{4} = \sqrt{\nu} a^{*}.
\end{align}
Here $N_{ex}$ the effective pump rate (number of atoms per cavity lifetime), and the parameter $\phi$ called the accumulated Rabi angle is proportional to $g$. Later we will consider that $\phi$ is an unknown parameter to be estimated. 

The operators $L_{i}$ can be interpreted as jump operators for different measurement processes:  detection of an output atom in the ground state or excited state ($L_{1}$ and $L_{2}$) and emission or absorption of a photon by the cavity
($L_{3}$ and $L_{4}$). In each case the cavity makes a jump up or down on the ladder of Fock states. Since both the atom-cavity interaction \eqref{eq.jaynes} and the cavity-bath interaction leave the commutative algebra of number operators invariant, we can restrict our attention to this classical dynamical system provided that the atoms are measured in the 
$\sigma_{z}$ basis. The cavity jumps are described by a birth-death (Markov) process with birth and death rates
\begin{eqnarray}
&&
q_{k,k+1}:= N_{ex} \sin(\phi \sqrt{k+1})^{2} + \nu (k+1) ,\quad k\geq 0\nonumber \\
&&
q_{k,k-1} := (\nu+1)k, \quad k\geq 1.\label{eq.birth-death}
\end{eqnarray}
In section \ref{sec.Fisher} we will come back to the birth-death process, in the context of estimating $\phi$.

\begin{figure}
\begin{center}
%\includegraphics[width=0.34\textwidth]{bistable1}  \quad  
%%{\label{fig:bistablegrid}
\includegraphics[width=0.7\textwidth]{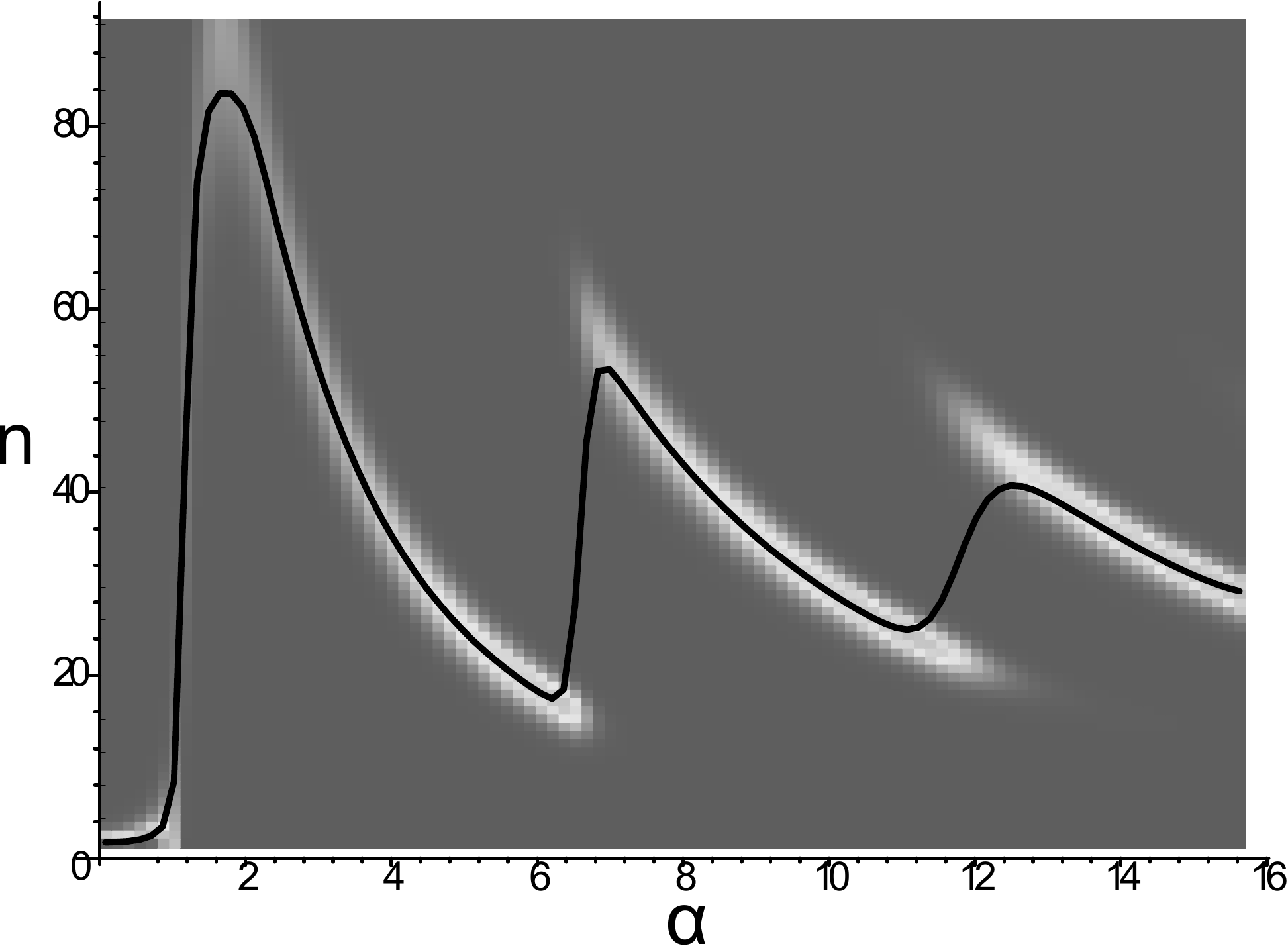}
\caption{The stationary state as function of $\alpha := \phi \sqrt{N_{ex}}$, for $\nu = \sqrt{1.15}, N_{ex} = 100$. 
%Left panel: stationary state for $\alpha = 3.8$ (dashed) and $\alpha = 7.45$ (solid). Right panel: 
The white patches represent the photon number distribution and the black line is the expected photon number.}
    \label{fig:stationary}
    \end{center}
\end{figure}

The Lindblad generator \eqref{eq.lindblad} has a unique stationary state ( i.e. $\mathcal{L}(\rho_{\mbox{s}})=0)$ which is diagonal in the Fock basis and has coefficients 
\begin{equation}\label{eq.stationary.state}
		\rho_{\mbox{s}}(n) = \rho_{\mbox{s}}(0) \prod_{k=1}^{n} \left( \frac{\nu}{\nu +1} + \frac{N_{ex}}{\nu +1} \frac{\sin^{2}(\phi \sqrt{k})}{k} \right)	.
	\end{equation}
This means that the cavity evolution is ergodic in the sense that in the long run any initial state converges to the stationary state. In figure \ref{fig:stationary} we illustrate the dependence on $\alpha:=\phi\sqrt{N_{ex}}$ of the stationary state. The notable features are the sharp change of the mean photon number at $\alpha\approx 1, 2\pi,4\pi...$, and the fact that the stationary state is bistable at these points with the exception of the first one. The bistability is accompanied by a significant narrowing of the spectral gap of $\mathcal{L}$. This behaviour is typical of a metastable Markov process where several `phases' are very weakly coupled to each other, and the process spends long periods in one phase before abruptly moving to another. 

An alternative perspective to these phenomena is offered by the counting process of the outgoing atoms. Since the rate  at which an atom exchanges an excitation with the cavity depends on the cavity state, the counting process carries information about the cavity dynamics, and in particular about the interaction parameter $\phi$.  The recent papers \cite{Garrahan2009,Garrahan2011} propose to analyse the stationary dynamics of the atom maser using the theories of {\it large deviations} and dynamical phase transitions. Instead of looking at the  `phases' of the stationary cavity state, the idea is to investigate  the long time properties of measurement trajectories and identify their {\it dynamical phases} 
i.e. ensembles of trajectories which have markedly different count rates in the long run, or `activities'. 
The large deviations approach raises important questions related to the existence of dynamical phase transitions 
which can be formulated in terms of the spectral properties of the `modified' generator $\mathcal{L}_{s}$ defined in section \ref{sec.Fisher}, and the Perron-Frobenius Theorem for infinite dimensional quantum Markov processes \cite{Guta&vanHorssen}. 
In this work we concentrate on the closely related, but distinct regime of {\it moderate deviations} characterised by the Central Limit Theorem, which is more relevant for statistical inference problems. For later purposes we introduce a unitary dilation of the master evolution which is defined by the unitary solution of the  following quantum stochastic differential equation 
\begin{equation}\label{eq.unitary.dilation}
d U(t) =  
\sum_{i=1}^{4} \left(L_{i} d A^{*}_{i,t} -  L_{i}^{*} d A_{i,t} - \frac{1}{2} L_{i}^{*} L_{i} dt\right) U(t).
\end{equation}
The pairs $(dA_{i,t},d A^{*}_{i,t})$ represent the increments of the creation and annihilation operators of 4 independent bosonic input channels, which couple with the cavity through the operators $L_{i}$. The master evolution can be recovered as usual by tracing out the bosonic environment which is initially in the vacuum state
$$
e^{t\mathcal{L} }(\rho) = {\rm Tr}_{env}(U(t) \left(\rho\otimes |\Omega\rangle\langle\Omega| \right)U(t)^{*})
$$
If $d\Gamma_{i,t}$ denote the increments of the four number operators of the input channels then the counting operators of the output are 
\begin{equation}\label{eq.lambdat}
\Lambda_{i,t} := U(t)^{*} \left(\mathbf{1}\otimes \Gamma_{i,t} \right)U(t),
\end{equation}
which provide the statistics of counting atoms in the ground state, excited state, emitted and absorbed photons. 
\section{Brief overview of classical and quantum statistics notions}
\label{sec.overview.statistics}
%%%%%%%%%%%%%%%%%%%%%%%%%%%%%%%%%
For reader's convenience we recall here some basic notions of classical and quantum parametric statistics which will be useful for interpreting the results of the next section.
%%%%%%%%%%%%%%%%%%%%%%%%%%%%%%%%%%%
\subsection{Estimation for independent identically distributed variables}
\label{sec.iid.classical}
%%%%%%%%%%%%%%%%%%%%%%%%%%%%%%%%%%%

A typical statistical problem is to estimate an unknown parameter $\theta = (\theta_{1},\dots,\theta_{k})\in \mathbb{R}^{k}$ given the data consisting of  independent, identically distributed samples 
$X_{1}, \dots ,X_{n}$ from a distribution $\mathbb{P}_{\theta}$ which depends on $\theta$.

An estimators $\hat{\theta}_{n}:=\hat{\theta}_{n}\left(X_{1},...,X_{n}\right)$ is called unbiased if 
$\mathbb{E}(\hat{\theta}_{n}) = \theta$ for all $\theta$. The Cram\'{e}r-Rao inequality gives 
a lower bound to the covariance matrix and mean square error of any unbiased estimator
\begin{eqnarray}
{\rm Cov} (\hat{\theta}_{n})= \mathbb{E}_{\theta}\left[ (\hat{\theta}_{n} -\theta )^{T}(\hat{\theta}_{n} -\theta ) \right]  &\geq& \frac{1}{n}I(\theta)^{-1} \label{eq.cr}\\
\mathbb{E}_{\theta} \left[ \| \hat{\theta}_{n} -\theta \|^{2} \right] &\geq& \frac{1}{n} \Tr \left(I(\theta)^{-1}\right).
\label{eq.mse}
\end{eqnarray}
The $k\times k$ positive matrix $I(\theta)$ is called the Fisher information matrix and can be computed in terms of the 
log-likelihood functions  $\ell_{\theta} := \log p_{\theta}$ where $p_{\theta}$ is the probability density of $\mathbb{P}_{\theta}$ with respect to some reference measure $\mu$:
$$
I (\theta)_{i,j}=
\mathbb{E}_{\theta}\left( \frac{\partial \ell_{\theta}}{\partial \theta_{i}} \frac{\partial \ell_{\theta}}{\partial \theta_{i}}\right)=
\int p_{\theta} (x )\frac{\partial \log p_{\theta}}{\partial\theta_{i}}\frac{\partial \log p_{\theta}}{\partial\theta_{j}}\mu\left(dx\right)
$$

The Cram\'{e}r-Rao bound is in general {\it not} achievable for a given $n$. However, what makes the Fisher information  important is the fact that the bound is {\it asymptotically achievable}. Furthermore, asymptotically optimal estimators (or efficient estimators)  are {\it asymptotically normal} in the sense that  
\begin{equation}\label{eq.asymptotic.normality}
\sqrt{n} (\hat{\theta}_{n} - \theta) \overset{\mathcal{L}}{\longrightarrow} N(0, I(\theta)^{-1})
\end{equation}
where the right side is a centred $k$-variate Gaussian distribution with covariance $ I(\theta)^{-1}$ and the convergence is in law for $n\to \infty$. Under certain regularity conditions, the maximum likelihood estimator
$$
\hat{\theta}_{n}: = \arg\max_{\theta^{\prime}} \prod_{i} p_{\theta^{\prime}}(X_{i}) 
$$
is efficient. The asymptotic normality of efficient estimators can be seen as a consequence of the more fundamental theory of {\it local asymptotic normality} (LAN) which states that the i.i.d. statistical model $\mathbb{P}_{\theta}^{n}$ can be `linearised' in a local neighbourhood of any point $\theta_{0}$ and approximated by a Gaussian model. Since the uncertainty in $\theta$ is of the order $n^{-1/2}$ we write 
$\theta:=\theta_{0}+ u/\sqrt{n}$ where $u$ is a local parameter which is considered unknown, while $\theta_{0}$ is fixed and known. Local asymptotic normality can be expressed as the (local) convergence of statistical models \cite{vanderVaart}
$$
\left\{ \mathbb{P}_{\theta_{0}+u/\sqrt{n}}^{n}\::\: u\in\mathbb{R}\right\} \rightarrow\left\{ N\left(u,I(\theta_{0})^{-1}\right)\::\: u\in\mathbb{R}\right\} 
$$
where the limit is approached as $n\to \infty$, and consists of a single sample from the normal distribution with unknown mean $u$ and known variance $I(\theta_{0})^{-1}$. In sections \ref{sec.Fisher}.(\ref{sec.q.fisher}) and \ref{sec.Fisher}.(\ref{sec.cl.fisher}) we will prove two versions of LAN, one for quantum states and one for a classical counting process.
\subsection{Quantum estimation with identical copies}
\label{sec.quantum.iid}
%%%%%%%%%%%%%%%%%%%%%%%%%%%%%
Consider now the problem of estimating $\theta\in\mathbb{R}^{k}$, given n identical and independent copies of a 
quantum state $\rho_{\theta}$. The quantum Cram\'e{r}-Rao bound \citep{Helstrom,Holevo,Braunstein&Caves} says that for any measurement on $\rho_{\theta}^{\otimes n}$ (including joint ones) and any unbiased estimator $\hat{\theta}_{n}$ constructed from the outcome of this measurement, the lower bound \eqref{eq.cr} holds with $I(\theta)$ replaced by the {\it quantum Fisher information matrix}
$$
F(\theta)_{i,j}=\Tr\left(\rho_{\theta} \mathcal{D}_{\theta,i}\circ\mathcal{D}_{\theta,j}\right)
$$
where $X\circ Y := \{X,Y\}/2$ and $\mathcal{D}_{\theta,i}$ are the self-adjoint operators defined by 
$$
\frac{\partial\rho_{\theta}}{\partial\theta_{i}}= \mathcal{D}_{\theta,i} \circ \rho_{\theta}.
$$

When $\theta$ is one dimensional, the quantum Cram\'e{r}-Rao bound is asymptotically achievable by the following two steps adaptive procedure. First, a small proportion $\tilde{n}\ll n$ of the systems is measured in a `standard' way and a rough estimator $\theta_{0}$ is constructed; in the second step, one measures $\mathcal{D}_{\theta_{0}}$ separately on each system to obtain results $D_{1},\dots, D_{n-\tilde{n}}$ and defines the efficient estimator 
$$
\hat{\theta}_{n} = \theta_{0} + \frac{1}{(n-\tilde{n})F(\theta_{0})} \sum_{i}D_{i} .
$$
However, for multi-dimensional parameters, the quantum Cram\'{e}r-Rao bound is not achievable even asymptotically, due to the fact that the operators $\mathcal{D}_{\theta,i}$ may not commute with each other and cannot be measured simultaneously. Moreover, unlike the classical case, there are several Cram\'{e}r-Rao bounds based on different notions of `Fisher information' \cite{Belavkin}. In this case it is more meaningful to search for asymptotically optimal estimators in the sense of optimising the risk given by mean square error \eqref{eq.mse}. In \cite{Hayashi&Matsumoto} it has been shown that for qubits, the asymptotically optimal risk is given by the so called Holevo bound \cite{Holevo}. For arbitrary dimensions, 
the achievability of the Holevo bound can be deduced from the theory of quantum local asymptotic normality developed in \cite{Kahn&Guta} and a discussion on this can be found in \cite{Guta&Kahn.proc}.

%%%%%%%%%%%%%%%%%%%%%%%%%%%%%
\subsection{Fisher information for classical Markov processes}
%%%%%%%%%%%%%%%%%%%%%%%%%%%%%

Often, the data we need to investigate is not a sequence of  i.i.d. variables but a stochastic process, e.g. a Markov 
process. A theory of efficient estimators and (local) asymptotic normality can be developed along the lines of the i.i.d. set-up, provided that the process is ergodic. We will describe the basic ingredients of a continuous time Markov process and write its Fisher information.

Let $I=\{ 1,...,m \} $ be a set of states, and  let $Q=[q_{ij}]$ be a $m\times m$ matrix of transition rates, with $q_{ij}\geq 0$ for 
$i\neq j$ and diagonal elements  $q_{ii}=-q_{i}:=-\sum_{j\neq i}q_{ij}$. 
The rate matrix is the generator of a continuous time Markov process, and the associated semigroup of transition operators is 
$$
P(t)=\exp(tQ).
$$

A continuous time stochastic process $\left(X_{t}\right)_{t\geq0}$ with state space $I$ is a Markov process with transition  semigroup $P(t)$ if
\[
\mathbb{P}\left(X_{t_{n+1}}=i_{n+1}|X_{t_{0}}=i_{0},...,X_{t_{n}}=i_{n}\right)=
p(t_{n+1}-t_{n})_{i_{n}i_{n+1}},
\]
 for all $n=0,1,2,...$, all times $0\leq t_{0}\leq...\leq t_{n+1}$,
and all states $i_{0},...,i_{n+1}$ where $p(t)_{ij}$
are the matrix elements of $P\left(t\right)$.

Let us denote by $J_{0},J_{1},...$ the times when the process jumps from one state to another, so that $J_{0}=0$ and $J_{n+1}=inf\left\{ t>J_{n}\,:\, X_{t}\neq X_{J_{n}}\right\} $.
The time between two jumps is called 'holding
time' and is defined by $S_{i}=J_{i}-J_{i-1}$.

A probability distribution $\pi= (\pi_{1},\dots, \pi_{m})$ over $I$ is stationary for the Markov process 
$\left(X_{t}\right)_{t\geq0}$ if it satisfies $\nu Q=0$ or equivalently $\pi P(t)=\pi$ at all $t$. 
If the transition matrix is irreducible then this distribution is unique and the process is called ergodic, in which case  any initial distribution $\mu$ converges to the stationary distribution 
$$
\lim_{t \to\infty} \mu P(t) = \pi.
$$
Suppose now that we observe the ergodic Markov process $X_{t}$ for $t\in [0,T]$, and that the rate matrix depends smoothly on some unknown parameter $\theta$ (which for simplicity we consider one dimensional), so that $q_{ij}= q^{\theta}_{ij}$. The asymptotic theory says that `good' estimators like maximum likelihood (under some regularity conditions) are asymptotically normal in the sense of \eqref{eq.asymptotic.normality}, with Fisher information given by
\begin{equation}\label{eq.fisher.markov}
I\left(\theta\right):=\sum_{i\neq j}\pi_{i}^{\theta} q_{ij}^{\theta} \left(V_{ij}^{\theta}\right)^{2}
\end{equation}
where 
\[
V_{ij}^{\theta}:=\frac{d}{d\theta}\log q_{ij}^{\theta}
\]
and $\pi^{\theta}$ is the stationary distribution at $\theta$.

%%%%%%%%%%%%%%%%%%%%%%%%%%%%%%%%%%
\section{Fisher informations for the atom maser}\label{sec.Fisher}
%%%%%%%%%%%%%%%%%%%%%%%%%%%%%%%%%%

In this section we return to the atom maser and investigate the problem of estimating the interaction parameter 
$\phi$, based on outcomes of measurements performed on the outgoing atoms. State and process tomography  are key enabling components in quantum engineering, and have become the focus of research at the intersection of quantum information theory and statistics. Our contribution is to go beyond the usual set-up of repeated measurements of identically prepared systems, or that of process tomography, and look at estimation in the quantum Markov set-up. The first step in this direction was made in \cite{Guta} which deals with asymptotics of system identification in a discrete time setting with finite dimensional systems. Here we extend these ideas to the atom maser, including the effect of the thermal bath. In the next subsections we consider several thought experiments in which counting measurements are performed in 
the output bosonic channels determined by the unitary coupling \eqref{eq.unitary.dilation}. While some of these scenarios are not meant to have a practical relevance, the point  is to analyse and compare the amount of  Fisher information carried by the various stochastic processes associated to the atom maser, as illustrated in Figures \ref{fig.Fisher} and \ref{fig.Fisher2}.

%%%%%%%%%%%%%%%%%%%%%%%%%%%%%
\subsection{Observing the cavity}
%%%%%%%%%%%%%%%%%%%%%%%%%%%%%

Consider first the scenario where all four channels are monitored by means of counting measurements. As already discussed in section \ref{sec.maser}, the conditional evolution of the cavity is described by the birth and death process consisting of jumps up and down the Fock ladder, with rates \eqref{eq.birth-death}.
%The key point to note is that if the cavity is initially in a Fock state, then it's conditional evolution will consist of random  jumps up or down the Fock ladder: a step up when an atom is detected in the ground state or a photon has been absorbed, and a step down when a photon is emitted. This is a special type of Markov process called birth and death process and its generator can be read off by restricting the Lindblad generator \eqref{eq.lindblad} to the commutative algebra of diagonal operators with respect to the Fock basis. 
%The only non-zero rates $q_{ij}$ are then
Note that when an atom is detected in the excited state, the cavity state remains unchanged, so the corresponding rate $N_{ex}\cos\left(\phi\sqrt{i+1}\right)^{2}$  does not appear in the birth and death rates.
%, but only in 
%\begin{eqnarray*}
%%q_{i,i+1} & = & N_{ex}\sin^{2}\left(\phi\sqrt{i+1}\right)+\nu\left(i+1\right)\\
%%q_{i,i-1} & =& \left(\nu+1\right)i\\
%q_{i,i}      & =& -\left(N_{ex}\cos^{2}\left(\phi\sqrt{i+1}\right)+\nu\left(i+1\right)+\left(\nu+1\right)i\right).
%\end{eqnarray*}
Later we will see that these atoms {\it do} carry Fisher information about the interaction parameter even if they do not modify the state of the cavity.

Since the cavity dynamics is Markovian, we can use \eqref{eq.fisher.markov} and the expression of the stationary state \eqref{eq.stationary.state} to compute the Fisher information of the stochastic process determined by the cavity state 
\[
I_{cav}\left(\phi\right)=\sum_{i=0}^{\infty}
\rho_{\mbox{s}}^{\phi}\left(i\right) 
\left(\frac{\left(q_{i,i+1}^{\phi}\right)^{'}}{q_{i,i+1}^{\phi}}\right)^{2}q_{i,i+1}^{\phi} 
%\sum_{i=1}^{\infty}\rho_{st}^{\phi}\left(i\right)  \left(\frac{\left(q_{i,i-1}^{\phi}\right)^{'}}{q_{i,i-1}^{\phi}}\right)^{2}q_{i,i-1}^{\phi} .
\]
We stress that this information refers to an observer who only has access to the cavity state, and cannot infer whether a jump up is due to exchanging an excitation with an atom or absorbing a photon from the bath.  Moreover, the observer does not get any information about atoms passing through the cavity without exchanging the excitation.

\begin{figure}[h]
\begin{center}
\includegraphics[scale=0.5]{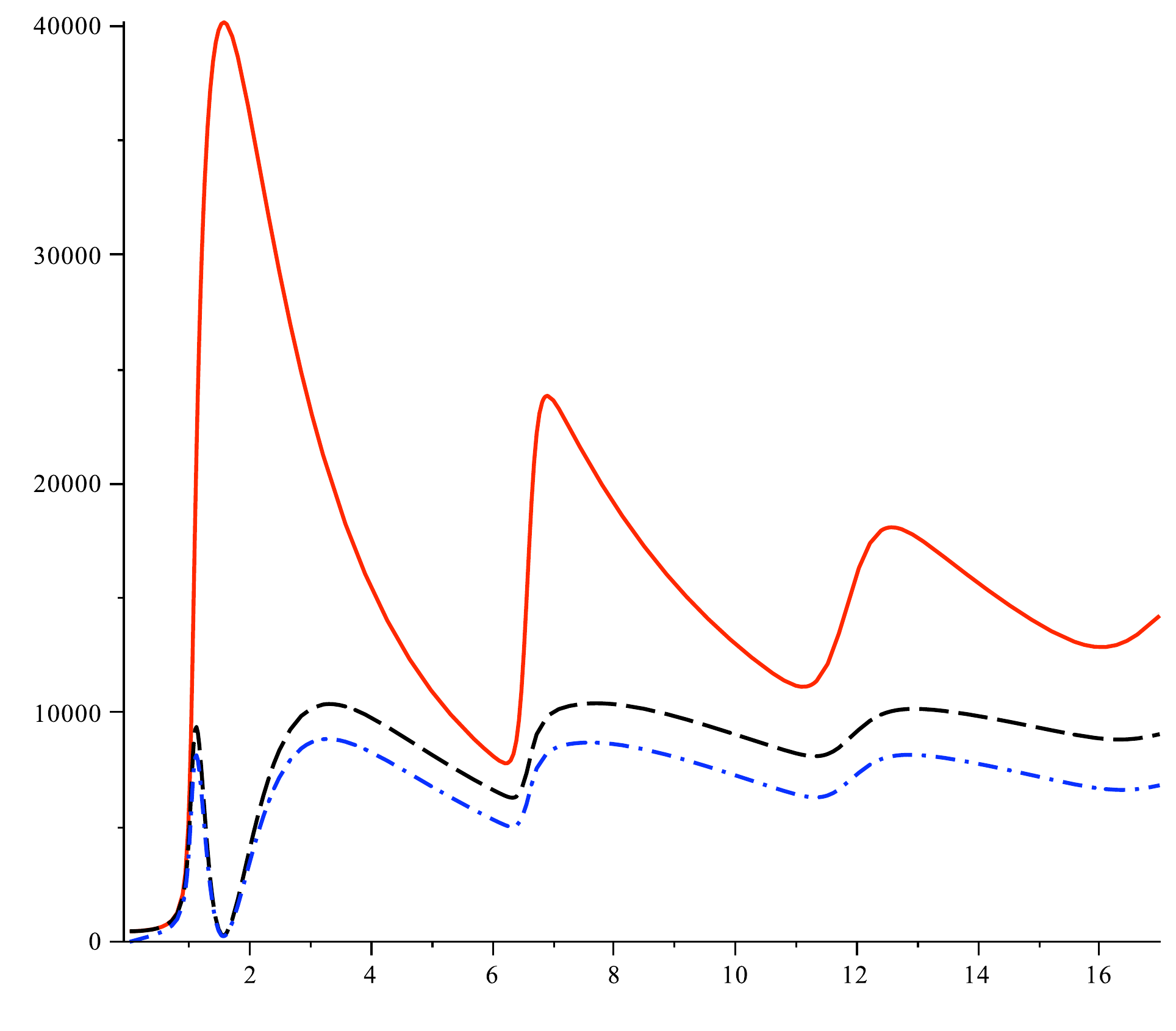}
\end{center}
\caption{(Asymptotic) Fisher informations for $\phi$ as function of $\alpha= \sqrt{N_{ex}}\phi$ in different scenarios (with 
$N_{ex}=100$ and $\nu=0.15$) : 
(a) when  the cavity is observed (blue dash-dot line) (b) when the type of the up jump is identified (black dash line) 
(c) when all four channels are monitored (cotinuous red line) (d) the quantum Fisher information of the output coincides with the classical information of (c).  
%(e) when we count the total number of ground state atoms in a time period (green dash dotted line)
}\label{fig.Fisher}
\end{figure}

The function $I_{cav}\left(\phi\right)$ is plotted as the dash-dot blue line in Figure \ref{fig.Fisher}.

\subsection{Observing the cavity and discriminating between jumps}

In the next step, we assume that besides monitoring the cavity, we are also able to discriminate between the two events producing a jump up, which in effect is equivalent to monitoring the emission and absorption from the bath and the atoms exiting in the ground state, but not those in the excited state.

But how do we model probabilistically the additional piece of information ? Let us fix a given trajectory of the cavity which has jumps up at times $t_{1},\dots, t_{l}$ from the Fock states with photon numbers $k_{1},\dots , k_{l}$. 
Conditional on this trajectory, the events  ``jump at $t_{i}$ is due to atom'' and its complement ``jump at $t_{i}$ is due to bath'' have probabilities 
$$
p_{a}^{i} =\frac{r^{k_{i}}_{a}}{r^{k_{i}}_{a}+ r^{k_{i}}_{b}},
%\frac{ N_{ex}\sin^{2}\left(\phi\sqrt{k_{i}+1}\right)}{ N_{ex}\sin^{2}\left(\phi\sqrt{k_{i}+1}\right)+\nu\left(k_{i}+1\right)}  ,
\qquad
p_{b}^{i} = \frac{r^{k_{i}}_{b}}{r^{k_{i}}_{a}+ r^{k_{i}}_{b}}
%\frac{\nu\left(k_{i}+1\right)}{ N_{ex}\sin^{2}\left(\phi\sqrt{k_{i}+1}\right)+\nu\left(k_{i}+1\right)} 
$$
where $r^{k}_{a}=N_{ex}\mbox{sin}\left(\phi\sqrt{k+1}\right)^{2} $ and $r^{k}_{b}= \nu\left(k+1\right)$ are the rates for atoms and bath jumps. This means that we can model the process by independently tossing a coin with probabilities $p_{a}^{i}$ and $p_{b}^{i}$ at each time $t_{i}$. For each toss the additional Fisher information is 
$$
I_{t_{i}} (\phi) = \left(\frac{dp_{a}^{i}}{d\phi}\right)^{2}\frac{1}{p_{a}^{i}( 1-p_{a}^{i})},
$$
and the information for the whole trajectory is obtained by summing over $i$. The (asymptotic) Fisher information of the process is obtained by  taking the time and stochastic averaging over trajectories, in the large time limit. Since in the 
long run the system is in the stationary state, the average number of  $k\to k+1$  jumps per unit of time is 
$\rho_{\mbox{s}}^{\phi}\left(k\right)q_{k,k+1}^{\phi}$, so the additional Fisher information provided by the jumps  is 
 \[
I_{up}\left(\phi\right)=\sum_{k=0}^{\infty}\rho_{\mbox{s}}^{\phi}\left(k\right)q_{k,k+1}^{\phi}\left(\frac{d p_{a}^{k}}{d\phi}\right)^{2}\frac{1}{p_{a}^{k}(1- p_{a}^{k})}.
\]
Therefore the Fisher information gained by following the cavity and discriminating between jumps is 
\[
I_{cav+up}\left(\phi\right)=I_{cav}\left(\phi\right)+I_{up}\left(\phi\right)
%=\sum_{i=0}^{\infty}\rho_{st}^{\phi}\left(i\right)\frac{1}{r_{1}^{i}}\left(q_{i,i+1}^{\phi}\right)^{'\,2}
\]
which is plotted as the black dash line in Figure \ref{fig.Fisher}.

%%%%%%%%%%%%%%%%%%%%%%%%%%%%%
\subsection{Observing all counting processes}
%%%%%%%%%%%%%%%%%%%%%%%%%%%%%

The next step is to incorporate the information contained in the detection of excited atoms, to obtain the full classical 
Fisher information of all four counting measurements. We will consider again a fixed cavity trajectory and compute the additional (conditional) Fisher information provided by the counts of excited atoms. During each holding time period 
$S_{i}= t_{i+1}-t_{i}$ the cavity is in the state $k_{i}$ and the excited atoms are described by a Poisson process with rate 
$$
r^{k_{i}}_{e}:=N_{ex}\cos\left(\phi\sqrt{k_{i}+1}\right)^{2}.
$$ 
Moreover the Poisson processes for different holding times are independent, so the conditional Fisher information is the sum of informations for each Poisson process. Now, for a Poisson process the {\it total} number of counts in a time interval is a sufficient statistic and the times of arrival can be neglected. Thus, we only need to compute the Fisher information of a Poisson {\it distribution} with mean $\lambda_{i}:=r^{k_{i}}_{e}S_{i}$ and add up over $i$. A short calculation shows that this is equal to 
$$
J_{i}=\left(\frac{d\lambda_{i}}{d\phi}\right)^{2}\frac{1}{\lambda_{i}} = 
\left(\frac{d {r^{k_{i}}_{e}}}{d\phi}\right)^{2}\frac{S_{i}}{r^{k_{i}}_{e}}. 
$$   
As before, it remains to add over all holding times and take average over trajectories, and average over a long period of time. This amounts to replacing $S_{i}$ by the stationary distribution $\rho_{\mbox{s}}(k_{i})$ which is the average time in the state $k_{i}$ per unit of time. The Fisher information is
$$
I_{exc}= \sum_{k=0}^{\infty} \rho_{\mbox{s}}(k) \frac{1}{r^{k}_{e}} \left(\frac{d {r^{k}_{e}}}{d\phi}\right)^{2}.
$$

We can now write down the total classical Fisher information of the four counting processes
\[
I_{tot}:= I_{exc}+ I_{up}+I_{cav}= 4N_{ex} \sum_{k=0}^{\infty}\rho_{\mbox{s}}(k)\left(k+1\right).
\]
where the last equality follows from a simple calculation based on the explicit expressions of the three terms. The total information $I_{tot}$ is plotted as continuous red line in Figure \ref{fig.Fisher}.

The last expression is surprisingly simple, and as we will see in the next section, it is equal to the 
{\it quantum} Fisher information of the atom maser output process, which is the maximum information extracted by {\it any} measurement!

%%%%%%%%%%%%%%%%%%%%%%%%%%%%%
\subsection{The quantum Fisher information of the atom maser}\label{sec.q.fisher}
%%%%%%%%%%%%%%%%%%%%%%%%%%%%%
Up to this point we considered the problem of estimating $\phi$  in several scenarios involving counting processes. We will now investigate the more general problem of estimating $\phi$ when arbitrary measurements are allowed. As discussed in section \ref{sec.overview.statistics}\ref{sec.quantum.iid}, the key statistical notions of Cram\'{e}r-Rao bounds, Fisher information and asymptotic normality can be extended to i.i.d. quantum statistical models, and can be used to find asymptotically optimal measurement strategies for parameter estimation problems. In \cite{Guta} these notions were extended to the non-i.i.d. framework of a quantum Markov chain with finite dimensional systems. Here we extend these results further to the atom maser, which is a continuous time Markov process with a infinite dimensional system. The general mathematical theory is developed in forthcoming paper \cite{Guta&Bouten} and we refer to \cite{Guta} more details on the physical and statistical interpretation of the results. 

Let $|\chi\rangle$ be the initial state of the cavity and $|\Omega\rangle$ the joint vacuum state of the bosonic fields. 
The joint (pure) state of the cavity and the four Bosonic channels at time $t$ is 
$$
|\psi_{\phi}(t)\rangle = U_{\phi}(t) \left( | \chi \rangle \otimes | \Omega\rangle  \right)
$$
where $U_{\phi}(t)$ is the unitary solution of the quantum stochastic differential equation \eqref{eq.unitary.dilation}. We emphasise that both the unitary and the state depend on the parameter $\phi$ and we would like to know what is the ultimate precision limit for the estimation of $\phi$ assuming that arbitrary measurements are available.

As argued in section \ref{sec.overview.statistics}(\ref{sec.iid.classical}), for asymptotics it suffices to understand the statistical model in a local neighbourhood of a given point, whose size is of the order of the statistical uncertainty, in this case  $t^{-1/2}$. For this we write 
$\phi= \phi_{0}+ u/\sqrt{t}$ and focus on the structure of the quantum statistical model with parameter $u\in \mathbb{R}$:
$$
|\psi(u,t)\rangle  := \left|\psi_{\phi_{0}+u/\sqrt{t}}(t)\right\rangle.
$$

Our main result is to show that this quantum model is {\it asymptotically Gaussian}, in the sense that this family of vectors converges to a family of coherent state of a continuous variables system, similarly to results obtained in \cite{Guta&Kahn,Guta&Janssens&Kahn,Guta&Jencova,Kahn&Guta} for identical copies of quantum states, and in \cite{Guta} for quantum Markov chains. More precisely
\begin{equation}\label{eq.asymptotic.coherent}
\lim_{t\rightarrow\infty}\left\langle\psi(u,t) | \psi(v,t)\right\rangle =\left\langle \sqrt{2F}\, v|\sqrt{2F}\, u\right\rangle =\mbox{e}^{-\left(u-v\right)^{2}/8F}
\end{equation}
where $|\sqrt{2F}\, u\rangle$ denotes a coherent state of a one mode continuous variables system, with displacement 
$\sqrt{2F}\, u$ along one axis, and $F=F(\phi_{0})$ is a constant which plays the role of quantum Fisher information  (per unit of time). The meaning of this result is that for large times, the state of the atom maser and environment is approximately Gaussian when seen from the perspective of parameter estimation, and by performing an appropriate measurement we can extract the maximum amount of information $F$. At the end of the following calculation we will find that $F=I_{tot}$, so the counting measurement is in fact optimal!  Recall however that the counting measurement involves the detection of emitted and absorbed photons which is experimentally unrealistic. However, the result is relevant as it puts an upper bound on {\it any} Fisher information that can be extracted by measurements on the output. To prove \eqref{eq.asymptotic.coherent} we express the inner product in terms of a (non completely positive) semigroup on the cavity space, by tracing over the atoms and bath 
\begin{equation}%\label{eq.asymptotic.coherent}
\left\langle\psi(u,t) | \psi(v,t)\right\rangle = 
\langle \phi|  e^{t\mathcal{L}_{u,v}} (\mathbf{1} )  |\phi\rangle
\end{equation}
where the generator $\mathcal{L}_{u,v}$ is 
\[
\mathcal{L}_{u,v}\left(X\right)=\sum_{i=1}^{4}\left( L_{i}^{u*}X  L_{i}^{v} -\frac{1}{2} L_{i}^{u*}L_{i}^{u}\, X-\frac{1}{2}X L_{i}^{v*}L_{i}^{v}\right)
\]
and $L_{i}^{u}= L_{i}(\phi_{0}+ u/\sqrt{t})$ are the operators appearing in the Lindblad generator \eqref{eq.lindblad}, where we emphasised the dependence on the local parameter. The proof of \eqref{eq.asymptotic.coherent} uses a second order perturbation result of Davies \cite{Davies} which will be discussed in more detail in the next section. Here we give the final result which says that the quantum Fisher information is proportional to the 
mean energy of the cavity in the stationary state, and is equal to the classical Fisher information $I_{tot}$ for the counting measurement : 
\[
F=4N_{ex}\sum_{k=0}^{\infty} \rho_{\mbox{s}} (k) (k+1).
\]

%%%%%%%%%%%%%%%%%%%%%%%%%%%%%
\subsection{Counting ground or excited state atoms}\label{sec.cl.fisher}
%%%%%%%%%%%%%%%%%%%%%%%%%%%%%

We now consider the scenario in which the estimation is based on the {\it total} number of ground state atoms 
$\Lambda_{1,t}$ defined in \eqref{eq.lambdat}, ignoring their arrival times. A similar argument can be applied to the excited state atoms. The generating function of $\Lambda_{t}$ can be computed from the unitary dilation \eqref{eq.unitary.dilation} which gives
\begin{equation} \label{eq:mgf}
\mathbb{E}\left(\exp \left( s \Lambda_{1,t} \right) \right) = \Tr \left( \rho_{0} e^{t \mathcal{L}_{s}} ({\bf 1}) \right)
	\end{equation}
where $\rho_{0}$ is the initial state of the cavity and $\mathcal{L}_{s}$ is the modified generator
\begin{equation} \label{eq:modgen}
		\mathcal{L}_{s}(\rho) = e^{s} L_{1} \rho L_{1}^{*}  - \frac{1}{2}\lbrace L_{1}^{*} L_{1}, \rho \rbrace+  \sum_{j\neq 1}\left( L_{j} \rho L_{j}^{*} - \frac{1}{2}\lbrace L_{j}^{*} L_{j}, \rho \rbrace \right).
\end{equation}
Note that $\mathcal{L}_{s}$ is the generator of a completely positive but not trace preserving semigroup.  We will analyse the {\it moderate} deviations of $\Lambda_{1,t}$ and show that it satisfies the Central Limit Theorem. In what concerns the estimation of $\phi$ we find an explicit expression of the Fisher information and establish asymptotic normality. The latter means that  
\begin{equation}\label{eq.asymptotic.normality.Lambda}
\tilde{\Lambda}_{1,t}:= \frac{1}{\sqrt{t}} (\Lambda_{1,t} - \mathbb{E}_{\phi_{0}}(\Lambda_{1,t})) \overset{\mathcal{L}}{\longrightarrow} N(\mu u, V)
\end{equation}
where the convergence holds as $t\to\infty$, with a fixed local parameter, i.e $\phi= \phi_{0}+u/\sqrt{t}$. In particular, for $u=0$ we recover the Central Limit Theorem for $\Lambda_{1,t}$. From \eqref{eq.asymptotic.normality.Lambda} we find that the estimator 
$$
\hat{\phi}_{t} := \phi_{0} + \frac{1}{\sqrt{t}} \tilde{\Lambda}_{1,t}/\mu 
$$
is efficient (as well as the maximum likelihood estimator), in the sense that its (rescaled) asymptotic variance $t {\rm Var}(\hat{\phi}_{t})$ is equal to the inverse of the Fisher information of the total counts of  ground state atoms
$$
I_{gr}= \mu^{2}/V.
$$ 
In the rest of the section we describe the main ideas involved in proving \eqref{eq.asymptotic.normality.Lambda} and give the expressions of $\mu$ and $V$. We first rewrite \eqref{eq.asymptotic.normality.Lambda}  in terms of the moment generating functions
\begin{equation}\label{eq.lan.lambda.genfct}
\varphi(s,t):=\mathbb{E}\left[\exp\left( i\frac{s}{\sqrt{t}} (\Lambda_{1,t}-\mathbb{E}_{\phi_{0}} (\Lambda_{1,t})) \right) \right]
\rightarrow\exp\left( i\mu u s -\frac{1}{2}s^{2}V\right).
\end{equation}
Using \eqref{eq:modgen} and \eqref{eq:mgf} with $s$ replaced by $s/\sqrt{t}$, the left side can be written as
$$
\varphi(s,t)= {\rm Tr}\left[\rho_{0} e^{t \mathcal{L}\left(\frac{s}{\sqrt{t}}, \frac{u}{\sqrt{t}}\right)} ({\bf 1}) \right] , \quad\text{with}\quad
\mathcal{L}\left(\frac{s}{\sqrt{t}}, \frac{u}{\sqrt{t}}\right) =\mathcal{L}_{\frac{s}{\sqrt{t}}} -\frac{1}{\sqrt{t}} \mathbb{E}_{\phi_{0}} \left(\frac{\Lambda_{t}}{t}\right),
$$ 
where $\rho_{0}$ is the initial state of the cavity.  The generator can be expanded in $t^{-1/2}$
$$
\mathcal{L}(s/\sqrt{t}, u/\sqrt{t}) 
= \mathcal{L}^{(0)}+\frac{1}{\sqrt{t}}\mathcal{L}^{(1)}+\frac{1}{t}\mathcal{L}^{(2)}+O\left(t^{-3/2}\right)
$$
and by applying the second order perturbation theorem  5.13 of \cite{Davies} we get  
\[
\lim_{t\to\infty} \varphi(s,t)=
\exp\left(
\Tr \left(\rho_{\mbox{s}} \,\mathcal{L}^{(2)} (\mathbf{1}) \right) -
\Tr \left(\rho_{\mbox{s}} \,\mathcal{L}^{(1)}\circ\tilde{\mathcal{L}}\circ\mathcal{L}^{(1)}(\mathbf{1}) \right)\right) 
\]
where $ \tilde{\mathcal{L}}$ is effectively the inverse of the restriction of $\mathcal{L}^{(0)}$ to the subspace of operators 
$X$ such that $\Tr (\rho_{\mbox{s}} X)=0$, which contains $\mathcal{L}^{(1)}(\mathbf{1})$. From the power expansion it can be seen that the the expression inside the last exponential is quadratic in $u,s$ which provides the formulas for $\mu$ and 
$V$ in \eqref{eq.lan.lambda.genfct}. The method outlined above is very general and can be applied to virtually any ergodic quantum Markov process. However in numerical computations we found that the fact $\mathcal{L}^{(0)}$ has a small spectral gap for certain values of $\phi_{0}$ may pose some difficulties in computing the inverse $\tilde{\mathcal{L}}$. An alternative method which we do not detail here is based on large deviation theory and shows that 
$$
\mu= \left.\frac{dr(s)}{ds}\right|_{s=0}, \qquad V= \left.\frac{d^{2}r(s)}{ds^{2}} \right|_{s=0}
$$
where $r(s)$ is the dominant eigenvalue of $\mathcal{L}_{s}$. Moreover, the coefficient $\mu$ can be computed in a more direct way as $\mu= d\Tr(\rho^{\phi}_{\mbox{s}}N ) /d\phi$ since for large times
$$
\mathbb{E}_{\phi}\left(\frac{\Lambda_{1, t} }{t} \right)=  \Tr(\rho^{\phi}_{\mbox{s}}N ) - \nu = \sum_{k=0}^{\infty}\rho^{\phi}_{\mbox{s}}(k) k -\nu
$$
which follows from an energy conservation argument in the stationary state.

\begin{figure}[h]
\begin{center}
\includegraphics[width=0.7\textwidth]{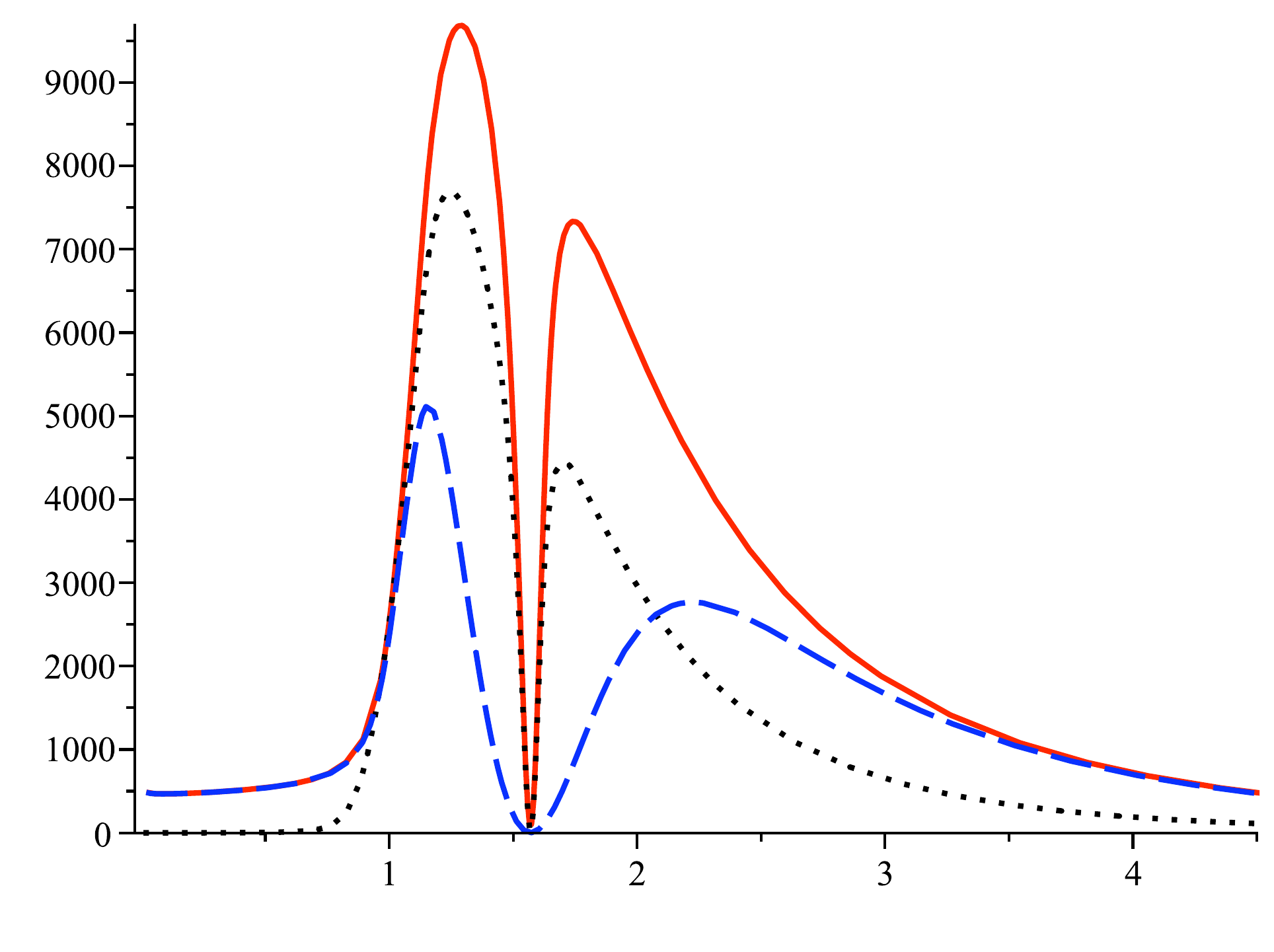}
\end{center}
\caption{The Fisher information for the total counts of ground state atoms (blue dash-dot line), excited state atoms 
(black dash line) and both counts together (red line) as function of $\alpha= \sqrt{N_{ex}}\phi $ at $N_{ex}=100$ and $\nu=0.15$.}\label{fig.Fisher2}
\end{figure}

A similar argument can be made for the total counts of the excited state atoms. 
The Fisher informations of both ground and excited state atoms are represented in  Figure \ref{fig.Fisher2}. The Fisher information for both counting processes together cane computed as well and is represented by the red line.
We note that the counts Fisher informations are comparable to those of the previous scenarios (see Figure \ref{fig.Fisher}) in the region $0\leq \alpha\leq 4$, but significantly smaller in the bistability regions. Also, they are equal to zero at $\phi\approx 0.16$ due to the fact that the derivative with $\phi$ of the mean atom number is zero at this point.

%%%%%%%%%%%%%%%%%%%%%%%%%%%%%
\section{Conclusions and outlook}
%%%%%%%%%%%%%%%%%%%%%%%%%%%%%

We have investigated the problem of estimating the Rabi frequency of the atom maser in the framework of asymptotic statistics. The Fisher informations of several classical counting processes were computed, together with the quantum Fisher information which is the upper bound of the classical information obtained from an arbitrary measurement. The latter was found to be equal to the $4N_{ex}\langle N+1\rangle_{\mbox{s}}$,  and is attained by the joint counting process of ground and excited atoms plus emitted and absorbed photons. However in the region of the first transition point we find that the Fisher information for both ground and excited total atom counts are equal to zero, while the quantum Fisher information is maximum. Even counting photons plus ground state atoms while ignoring the excited atoms, does not give a significant amount of information. It would be interesting to see whether estimation precision at this point can be improved by taking into account the full atom counts trajectories. Although maximum likelihood can be applied to these processes, perhaps in conjunction with Bayesian estimation and state filtering methods, this may be rather expensive in terms of  computational time.  An alternative is to use other estimation methods which are not likelihood based, e.g. approximate Bayesian computation methods. Another future direction  is to explore the relation between the moderate deviations regime which we have analysed here, and the large deviations regime which is relevant for the study of dynamical phase transition \cite{Garrahan2009,Garrahan2011}.  Ultimately the goal is to design measurements  which optimise the statistical performance of the estimation, in the spirit of Wiseman's adaptive phase estimation protocol  \cite{Wiseman} and to explore the connections with control theory, e.g. in the frame of adaptive control. Two papers detailing the proofs of the asymptotic normality results in a general Markov set-up \cite{Guta&Bouten} and the large deviations perspective \cite{Guta&vanHorssen} are in preparation.

%\bibliographystyle{plain}

%\bibliography{bibliography,bibsud,tesibib}
%\bibliographystyle{utphys}

%\appendix
%\section{Appendix}

\end{document}